\documentclass[article,10 pt,conference]{ieeeconf}  

\IEEEoverridecommandlockouts                              

\usepackage{epsfig} 
\usepackage{color} 
\usepackage{amsmath} 
\usepackage{amssymb}  
\usepackage{hyperref}
\newtheorem{theorem}{Theorem}

\usepackage{enumerate}
\newtheorem{proposition}{Proposition}

\newtheorem{remm}{Remark}
\newenvironment{remark}{\begin{remm}\rm }{\hfill \hspace*{1pt} \hfill $\circ$\end{remm}}

\newcommand\hspa{\hspace{-.5cm}}
\newcommand\Ltwo{\ensuremath{{\mathcal L}_2}}
\newcommand\real{\ensuremath{{\mathbb R}}}

\DeclareMathOperator{\sat}{sat}
\DeclareMathOperator{\dz}{dz}
\DeclareMathOperator{\He}{He}

\def\Q{\mathbb{Q}}
\def\S{\mathcal{S}}
\def\B{\mathrm{B}}

\newcommand{\ms}{{\textbf{q}}}
\newcommand{\ped}[1]{{_{\mathrm{#1}}}}

\def\tRO{\theta\ped{RO}}
\def\tSO{\theta\ped{SO}}
\def\tROC{\theta\ped{RwC}}
\def\tSOC{\theta\ped{SwC}}
\def\tSSOC{\theta\ped{SSwC}}
\def\V{\mbox{\rm Viol}}
\def\R{\mbox{\rm Rel}}

\title{\LARGE \bf
Scenario optimization with certificates\\ and applications to anti-windup design
}

\author{Simone Formentin, Fabrizio Dabbene, Roberto Tempo, Luca Zaccarian, Sergio M. Savaresi
\thanks{This work was supported in part by the ANR project LimICoS contract number 12 BS03 005 01, and by HYCON2 Network of Excellence ``Highly-Complex and Networked
Control Systems", grant agreement 257462.}
\thanks{Simone Formentin and Sergio M. Savaresi are with Dipartimento di Elettronica, Informazione e Bioingegneria, Politecnico di Milano, Piazza Leonardo da Vinci 32, 20133 Milano, Italy.}%
\thanks{Luca Zaccarian is with CNRS, LAAS, 7 avenue du colonel Roche, F-31400 Toulouse, France,
Univ. de Toulouse, LAAS, F-31400 Toulouse, France, and Dip. di Ingegneria Industriale, University of Trento, Italy.}%
\thanks{Fabrizio Dabbene and Roberto Tempo are with CNR-IEIIT, Corso Duca degli Abruzzi 123, Torino, Italy.}%
\thanks{E-mail to: {\tt\small simone.formentin@polimi.it}.}
}

\begin{document}

\maketitle
\thispagestyle{empty}
\pagestyle{empty}

\begin{abstract}
In this paper, we introduce a significant extension, called \textit{scenario with certificates} (SwC), of the so-called scenario approach for uncertain optimization problems. This extension is motivated by the observation that in many control problems only \textit{some} of the optimization variables are used in the design phase, while the other variables play the role of \textit{certificates}. Examples are all those control problems that can be reformulated in terms of linear matrix inequalities involving parameter-dependent Lyapunov functions. These control problems include static anti-windup compensator design for uncertain linear systems with input saturation, where the goal is the minimization of the nonlinear gain from an exogenous input to a performance output.
The main contribution of this paper is to show that randomization is a useful tool, specifically for anti-windup design, to make the overall approach less conservative compared to its robust counterpart. In particular, we demonstrate that the scenario with certificates reformulation is appealing because it provides a way to implicitly design the parameter-dependent Lyapunov functions.
Finally, to further reduce the computational cost of this one-shot approach, we present a sequential randomized algorithm for iteratively solving this problem.
\end{abstract}

\section{INTRODUCTION}

Randomized and probabilistic methods for control received a growing attention in the systems and control community in recent years \cite{TeCaDa:13}. These methods deal with the design of controllers for systems affected by possibly nonlinear, structured and unstructured uncertainties. One of the key features of these methods is to break the curse of dimensionality, \textit{i.e.}, uncertainty is ``lifted'' and the resulting controller satisfies a given performance for "almost" all uncertainty realizations. In other words, in this framework, we accept a ``small'' risk of performance violation.

One of the successful methods that have been developed in the area of randomized and probabilistic methods is the so-called scenario approach, which provides an effective tool for solving control problems formulated in terms of robust optimization \cite{CalCam:06tac}. In this case, the sample complexity, which is the number of random samples that should be drawn according to a given probabilistic distribution, is derived a priori, and it depends only on the number of design parameters $n_{\theta}$, and probabilistic parameters
called accuracy $\epsilon$ and confidence $\delta$. In particular, it has been shown that $n_{\theta}$, and $1/\epsilon$ enter linearly into the sample complexity and $\delta$ in a logarithmic fashion. Therefore, we can conclude that $\delta$ is computationally cheap, while accuracy is more expensive, and for very tiny values of $\epsilon$, the sample complexity becomes very large. For this reason, sequential approaches have been developed \cite{CDTV:13}. In this case, at each iteration of the algorithm, the idea is to construct (temporarily) a controller, whose performance is then validated through a Monte Carlo approach. If the controller does not enjoy the required probabilistic performance specification, a new controller is designed based on new sample extractions. At each step of the sequence, a reduced-size scenario problem is solved. This method is more effective in practical applications where the standard scenario approach fails because of the large sample complexity, but has other drawbacks including the fact that the sample complexity cannot be determined a priori.

The classical scenario approach and its sequential version may be applied to specific control problems such as designing a common quadratic Lyapunov function. In these cases, however, the fact that a single common Lyapunov function should hold for all possible uncertainties leads to an overly conservative design. The same drawback is known in classical robust control, where the design of a common quadratic Lyapunov function requires an exponential number of computations \cite{ATRC:08,CalDab:08b}. For these reasons, parameterized Lyapunov functions have been developed and used in many control problems subject to uncertainty \cite[Chapter 19.4]{Barmish:94}.

In the first part of the paper (Section \ref{sec:SwC_big}), we develop a new framework, denoted as scenario with certificates, which is very effective in dealing with parameter-dependent Lyapunov functions. This framework continues the research originally proposed in \cite{Oishi:07a} for feasibility problems in the context of randomized methods. The main idea in this approach is to distinguish between \textit{design} variables $\theta$ and \textit{certificates} $\xi$ and has the advantage, compared to classical robust methods, that no explicit parameterization (linear or nonlinear) of the Lyapunov functions is required. In other words, the method is based on a ``hidden" parameterization of the Lyapunov functions, and has the clear advantage compared to the methods based on the design of common Lyapunov functions to reduce the conservatism.

In the second part of the paper (Section \ref{ref:aw} and \ref{ref:example}), we show the application of the scenario with certificates approach to the design of anti-windup schemes. We recall that robust approaches for these schemes have been developed by proposing several successful solutions, see, \textit{e.g.}, \cite{MarcosTAC07,TurnerTAC07,Sofrony07,4380501,Post07}.

In \cite{ForSaZacDab:13}, the classical scenario optimization approach has been applied to anti-windup schemes in order to obtain probabilistic guarantees for the \Ltwo\ gain of the uncertain systems, without enforcing any constraint on the uncertainty set. However, in that formulation, both the certificates and the design variables were treated as optimization variables over the whole operating region, thus leading to a conservative solution, and sometimes - to unfeasibility. With the proposed SwC optimization approach, we show that robust compensators can be obtained without enforcing any additional constraints on the certificates. Moreover, by computing the (probabilistic) worst case \Ltwo\ gain with the SwC version of the analysis problem, we can show that the estimate of the worst case \Ltwo\ gain becomes much more accurate. This is demonstrated using the benchamrk example introduced in \cite{ZackAWbook11} (Section \ref{ref:example}) to show the performance of static anti-windup augmentation.
The paper is ended by some concluding remarks.

\section{SCENARIO WITH CERTIFICATES}\label{sec:SwC_big}

In this section, we briefly recall the scenario approach for dealing with convex optimization problems in the presence of uncertainty, and subsequently introduce a novel framework that we name scenario with certificates.

\subsection{The scenario approach}
The so-called scenario approach has been developed, \cite{CalCam:06tac} \cite{CaGAPa:09}, to deal with robust convex optimization problems of the form 
\begin{align}
\tRO  =  \arg &\displaystyle \min_{\theta\in\Theta} c^T \theta     \tag{RO}\label{eq:robust_opt} \\
                  &    \text{s.t. }   f(\theta,q) \leq 0,    \    \forall q  \in \Q,\nonumber
\end{align}
where, for given $q$ within the uncertainty set $\Q$, $f(\theta,q)$ are convex functions of the optimization variable $\theta\in\Theta$, and the domain $\Theta$ is a convex and compact set in $\real^{n_{\theta}}$.

Following the probabilistic approach discussed for instance in \cite{TeCaDa:13,CaDaTe:11} a probabilistic description of the uncertainty is considered over $\Q$, that is we formally assume that $q$ is a random variable with given probability distribution with support $\Q$. Such a probability distribution may describe the likelihood of each occurrence of the uncertainty or a user-defined weight for all possible uncertain situations. Then, $N$ independent identically distributed (iid) samples $q^{(1)}, \ldots , q^{(N)}$ are extracted according to the probability distribution of the uncertainty over $\Q$, forming the multisample
\[
\ms\doteq\left\{q^{(1)}, \ldots , q^{(N)}\right\}.
\]
These samples are used to construct the following scenario optimization (SO) problem,
based on $N$ instances (scenarios) of the uncertain constraints 
\begin{align}
         \tSO    =   \arg &\displaystyle\min_{\theta\in\Theta} c^T \theta    \tag{SO}\label{eq:scenario_opt} \\
                    & \text{s.t. }     f(\theta,q ^{(i)}) \leq 0, \  i=1,\ldots,N. \nonumber
\end{align}
The problem \eqref{eq:scenario_opt} can be seen as a probabilistic relaxation of the problem \eqref{eq:robust_opt}, since it deals only with a subset of the constraints considered in \eqref{eq:robust_opt}, according to the probability distribution of the uncertainty. However, under rather mild assumptions on problem \eqref{eq:robust_opt}, by suitably choosing~$N$, this approximation may in practice become negligible. Specifically, $N$ can be selected depending on the level of ``risk'' of constraint violation that the
user is willing to accept. Then, we define the \textit{violation probability} of the design $\theta$ as
\begin{equation}
\label{eq:viol}
\V(\theta)\doteq\Pr\left\{q\in\Q \,:\,f(\theta,q)>0\right\}.
\end{equation}
Similarly,  the \textit{reliability} of the design $\theta$ is given by
\[
\R(\theta)\doteq1-\V(\theta).
\]
Then the following result has been proven in \cite{CamGar:08}.
\begin{proposition}
\label{prop:scenario}
\cite{CamGar:08} Assume that, for any multisample extraction, problem (\ref{eq:scenario_opt}) is feasible and attains a unique optimal solution. Then,  given an accuracy level $\epsilon\in(0,1)$, the solution $\tSO$ of problem \eqref{eq:scenario_opt} satisfies
\begin{equation}\label{eq:binom}
\Pr\left\{\V(\tSO)>\epsilon\right\} \le \B(N,\epsilon,n_{\theta}),
\end{equation}
where
\begin{equation}\label{eq:binom2}
\B(N,\epsilon,n_{\theta}) \doteq \sum_{k=0}^{n_{\theta}-1}\binom{N}{k}\epsilon^{k}(1-\epsilon)^{(N-k)}.
\end{equation}
\end{proposition}

\smallskip

We note that non-uniqueness of the optimal solution can be circumvented by imposing additional ``tie-break'' rules in the problem, see, \textit{e.g.}, Appendix A of \cite{CalCam:06tac}.
Also, in \cite{Calafiore:10siopt} it is shown that the feasibility assumption can be removed at the expense of substituting $n_{\theta}-1$ with $n_{\theta}$ in $\B(N,\epsilon,n_{\theta})$.

From Equation (\ref{eq:binom}), explicit bounds on the number of samples necessary to guarantee the  ``goodness'' of the solution have been derived. The bound provided in \cite{AlTeLu:10} shows that, if, for given $\epsilon,\delta\in(0,1)$, the sample complexity $N$ is chosen  to satisfy the  bound
\begin{equation}\label{eq:N}
N\geq \frac{\mathrm{e}}{\epsilon(\mathrm{e}-1)}\left(\ln\frac{1}{\delta}+n_{\theta}-1\right)
\end{equation} 
(where $e$ denotes the Euler number), then the solution $\tSO$ of problem \eqref{eq:scenario_opt} satisfies $\V(\tSO)\le\epsilon$ with probability $1-\delta$. This bound shows that problem (SO) exhibits linear dependence in $1/\epsilon$ and $n_{\theta}$, and logarithmic dependence on $1/\delta$. Note however that, from a practical viewpoint, it is always preferable to numerically solve the one dimensional problem of finding the smallest $N$ such that $\B(N,\epsilon,n_{\theta})\le\delta$.

\smallskip

\subsection{Scenario with certificates}
The classical scenario approach previously discussed deals with uncertain optimization problems where  all  variables $\theta$ are to be designed. 
On the other hand, following the approach introduced in \cite{Oishi:07a} for feasibility problems, in the design with certificates approach we distinguish between \textit{design} variables $\theta$ and \textit{certificates} $\xi$. In particular, we consider now a function $f(\theta,\xi,q)$, jointly convex in $\theta\in\Theta$ and $\xi\in\Xi$ for given $q\in\Q$, and construct the robust optimization problem with certificates 
\begin{align}
\tROC  =   \arg &\displaystyle\min_{\theta} c^T \theta \tag{RwC}\label{eq:certificates_opt}\\
                  &  \text{s.t. }      \theta \in \S(q),    \    \forall q  \in \Q, \nonumber
\end{align}
where the set $\S(q)$ is defined as
\[
\S(q)\doteq
\left\{\theta\in\Theta | \ \exists \xi\in\Xi
\text{ satisfying } f(\theta,\xi,q)\le 0
\right\}.
\]
 
\begin{remark}[Convexity of problem \eqref{eq:certificates_opt}]
Note that, for any given $q$, the set $\S(q)$ is convex: to see this, consider $\theta_{1},\theta_{2}\in\S(q)$. Then, there exist $\xi_{1},\xi_{2}$ such that $f(\theta_{1},\xi_{1},q)\le 0$ and $f(\theta_{2},\xi_{2},q)\le 0$. Consider now  $\theta_{\lambda}\doteq \lambda \theta_{1}+(1-\lambda)\theta_{2}$, with $\lambda\in[0,\,1]$, and let $\xi_{\lambda}=\lambda \xi_{1}+(1-\lambda)\xi_{2}$. Then,  from convexity of $f$ with respect to both $\theta$ and $\xi$ it immediately follows
\[
f(\theta_{\lambda},\xi_{\lambda},q)\le \lambda f(\theta_{1},\xi_{1},q) + (1-\lambda)f(\theta_{2},\xi_{2},q)\le 0,
\]
hence $\theta_{\lambda}\in\S(q)$, which proves convexity.
\end{remark}

We propose to approximate problem \eqref{eq:certificates_opt} introducing the following \textit{scenario optimization  with certificates} problem, based again on a multisample extraction
\begin{align}
         \tSOC    =   \arg&\displaystyle \min_{\theta,\xi_{1},\ldots,\xi_{N}} c^T \theta  \tag{SwC}\label{eq:scenario_cert}\\
                    &  \text{s.t. }  f(\theta,\xi_{i},q^{(i)}) \leq 0, \  i=1,\ldots,N.\nonumber
\end{align}
Note that, contrary to problem \eqref{eq:scenario_opt}, in this case a new certificate variable $\xi_{i}$ is created for every sample $q^{(i)}$, $i=1,\ldots,N$. To analyze the properties of the solution $\tSOC$, we note that, in the case of SwC, the reliability and violation probabilities of design $\theta$ rewrite
\begin{eqnarray*}
\R(\theta)&=&
\Pr\Bigl\{q\in\Q | \exists \xi\in\Xi \text{ satisfying } f(\theta,\xi,q)\le 0 \Bigr\},\\
\V(\theta)&=&
\Pr\Bigl\{\exists q\in\Q  | \nexists \xi\in\Xi \text{ satisfying } f(\theta,\xi,q)\le 0 \Bigr\}.
\end{eqnarray*}

We now state the main result regarding the scenario optimization with certificates.
\begin{theorem}
\label{them:SwC}
Assume that, for any multisample extraction, problem \eqref{eq:scenario_cert} is feasible and attains a unique optimal solution.
Then, given an accuracy level $\epsilon\in(0,1)$, the solution $\tSOC$ of problem \eqref{eq:scenario_cert} satisfies
\begin{equation}\label{eq:binom_reprise}
\Pr\left\{\V(\tSOC)>\epsilon\right\}\le\B(N,\epsilon,n_{\theta}).
\end{equation}
\end{theorem}
\smallskip

\proof
Observe that the condition $\theta\in\S(q)$ is equivalent to requiring 
\[
f_{\xi}(\theta,q)\doteq\inf_{\xi\in\real^{n_{\xi}}}
f(\theta,\xi,q)\le 0,
\]
so that problem \eqref{eq:certificates_opt} is equivalent to
\begin{eqnarray}\label{eq:certificates_opt2}
 && \min_{\theta} c^T \theta \\
 & & \mbox{s.t. }   f_{\xi}(\theta,q)\le 0 \quad    \forall q  \in \Q.\nonumber
\end{eqnarray}
Note that, from joint convexity of $f$ with respect to $\theta,\xi$ for given $q$, it follows that for given $q$ the function $f_{\xi}(\theta,q)$ is convex in $\theta$, see for instance \cite[p.\ 113]{BoyVan:04}. Hence, problem (10) is a robust convex optimization problem, for which we can construct the following scenario counterpart
\begin{eqnarray}
&& \min_{\theta} c^T \theta,  \label{eq:scen2}\\
& & \mbox{s.t. } \min_{\xi_{i}\in\real^{n_{\xi}}} f(\theta,\xi_{i},q^{(i)}) \le 0, \, i=1,\ldots,N, \nonumber
\end{eqnarray}
where the subscript $i$ for the variables $\xi_{i}$ highlights that the different minimization problems are independent. Finally, we note that \eqref{eq:scen2} immediately rewrites as problem \eqref{eq:scenario_cert}.

\subsection{Sequential randomized algorithm for SwC}
\label{sec:SwC}

When $\epsilon$ is small and $n_{\theta}$ is large, the smallest value of $N$ satisfying (\ref{eq:N}) is typically very large. This fact makes the computational burden for SwC prohibitive for real world applications. In this section, we  present a sequential randomized algorithm that alleviates the computational load of solving the SwC problem in one-shot. The algorithm is a minor modification of \cite[Algorithm 1]{CDTV:13}, and it is based on separate design and validation steps. The design step requires the solution of a reduced-size SwC problem. In the validation step, contrary to \cite{CDTV:13} where only functional evaluations are required, a number of feasibility problems (one for each sample extraction) need to be solved. However, it should be pointed out that these problems are of small size, and can be solved independently, and hence parallelized.
The sequential procedure  is presented in Algorithm 1, and its theoretical properties are  stated in the subsequent theorem. The proof is omitted because it follows the same lines of \cite[Theorem 1]{CDTV:13}.\\

\noindent \textit{Sequential Algorithm for SwC}
\begin{enumerate} 
  \item \textsc{Initialization}\newline 
	set the iteration counter $k=0$. Choose the desired probabilistic levels $\epsilon$, $\delta$ and the desired number of iterations $k_t>1$
  \item \textsc{Update}\label{item:update}\newline
  set $k=k+1$ and $N_k\ge N\frac{k}{k_t}$ where $N$ is the smallest integer s.t. 
  $\B(N,\epsilon,n_{\theta})\leq\delta/2$
  \item \textsc{Design}
  \begin{itemize}
    \item draw $N_k$ iid (design) samples\\ $\ms_d=\{q_d^{(1)},\ldots, q_d^{(N_k)}\}\in\mathbb{Q}$
    \item solve the following \textit{reduced-size SwC problem}
\begin{eqnarray}\label{eq:SwC alg1 design}
        & \hat{\theta}_{N_k}    =&\!\!\!\!   \arg\displaystyle \min_{\theta,\xi_{1},\ldots,\xi_{N}} c^T \theta,       \\
          &          &\!\!\!\!\mbox{s.t. }     f(\theta,\xi_{i},q_{d}^{(i)}) \leq 0, \  i=1,\ldots,N.\nonumber
\end{eqnarray}
    \item \textbf{if} the last iteration is reached $(k=k_t)$, \textbf{return} $\tSSOC=\hat{\theta}_{N_k}$
  \end{itemize}
  \item \textsc{Validation}
  \begin{itemize}
    \item set $M_{k}$ according to \eqref{eq:sample bound Mk}
    \item draw iid (validation) samples $\ms_v=\{q_v^{(1)},\ldots, q_v^{(M_k)}\}\in\mathbb{Q}$ 
    \item \textbf{for} $j=1$ \textbf{to} $M_{k}$
  \begin{itemize}
    \item \textbf{if} the validation problem
\begin{eqnarray}\label{eq:SwC alg1 valid}
         &&   \text{find } \xi_{j} \mbox{ such that}     \nonumber \\
                    &   &  f(\hat{\theta}_{N_k},\xi_{j},q_{v}^{(j)}) \leq 0    
\end{eqnarray}
        is unfeasible  \textbf{goto} step (\ref{item:update}).
\end{itemize}
\item 
\textbf{return} $\tSSOC=\hat{\theta}_{N_k}$.
  \end{itemize}
\end{enumerate}

\begin{theorem}\label{theo:property of algorithm 1}
Assume that, for any multisample extraction, problem \eqref{eq:SwC alg1 design} is feasible and attains a unique optimal solution. Then,  given  accuracy level $\epsilon\in(0,1)$ and confidence level $\delta\in(0,1)$, let
    \begin{equation}\label{eq:sample bound Mk}
    M_k\ge\frac{\alpha\ln k+\ln \left(\mathcal{H}_{k_t-1}(\alpha)\right)+\ln\frac{2}{\delta}}{\ln\left(\frac{1}{1-\epsilon}\right)}
    \end{equation}
where $\mathcal{H}_{k_t-1}(\alpha)=\sum_{j=1}^{k_t-1}j^{-\alpha}$, with $\alpha>0$, is a finite hyperharmonic series.
Then, the probability that at iteration $k$  Algorithm 1 returns a solution $\tSSOC$
with violation greater than $\epsilon$ is at most $\delta$, \textit{i.e.},
\begin{equation}\label{eq:thr2}
\Pr\left\{\V(\tSSOC)>\epsilon\right\}\le\delta.
\end{equation}
\end{theorem}

In the second part of this paper, we introduce the problem of robust \Ltwo\ gain minimization for linear anti-windup systems. The SwC approach appears to be very suited for such a design problem, for several reasons: i) the nominal design can be formulated in terms of linear matrix inequalities (LMIs), ii) the uncertainty set can in principle be of any size and shape, and iii) the optimization variables can be clearly divided in design variables for the anti-windup augmentation and certificates for stability and performance guarantees.
\section{ANTI-WINDUP COMPENSATOR DESIGN}\label{ref:aw}
Consider the linear uncertain continuous-time plant with input saturation
\begin{eqnarray}
\label{eq:P}
\dot{x}_p         & =  &  A_p(q) x_p       +  B_{p,u}(q)\sigma   +  B_{p,w}(q)w \nonumber \\
    y             & =  &  C_{p,y}(q) x_p   +  D_{p,yu}(q)\sigma  +  D_{p,yw}(q)w \\
    z             & =  &  C_{p,z}(q) x_p   +  D_{p,zu}(q)\sigma  +  D_{p,zw}(q)w, \nonumber
\end{eqnarray}
where $x_p$ is the plant state, $\sigma$ is the control input, $w$ is an external input (possibly comprising references and disturbances), $z$ is the performance output, $y$ is the measured output and $q$ denotes a random uncertainty within the set $\Q$. We denote by $\bar q\in\Q$ the nominal value of the uncertain parameters.

As customary with linear anti-windup design \cite{ZackAWbook11}, we assume that a linear controller has been designed, based on the nominal system, in order to induce suitable nominal closed-loop properties when interconnected to plant (\ref{eq:P})
\begin{eqnarray}
\label{eq:C}
\begin{array}{rcl}
\dot{x}_c   & =  &  A_c x_c       +  B_{c,y}y   +  B_{c,w}w + v_1  \\
    u       & =  &  C_{c} x_c +  D_{c,y}y +  D_{c,w}w + v_2,
\end{array}
\end{eqnarray}
where $x_c$ is the controller state, $w$ typically comprises references (but may also contain disturbances), $u$ is the controller output and $v= [v_1^T \ v_2^T]^T$ is an extra input available for anti-windup action. Controller (\ref{eq:C}) is typically designed in such a way that the so-called {\em unconstrained closed-loop system} given by (\ref{eq:P}), (\ref{eq:C}), $\sigma=u$, $v=0$ is asymptotically stable and satisfies some performance requirements. 

Consider now the (physically more reasonable) {\em saturated interconnection} $\sigma=\sat(u)$, where the $k^{th}$ entry of $\sigma$ is $\sat_k (u_k) = \max(\min(\bar u_k,u_k),-\bar u_k)$, denoting the $k^{th}$ input by $u_k$. When the input saturates, the closed loop system composed by the feedback loop between (\ref{eq:P}) and (\ref{eq:C}) is no longer linear and may exhibit undesirable behavior, usually called {\em controller windup}. Then, one may wish to use the free input $v$ to design a suitable {\em static anti-windup compensator} of the form 
\begin{eqnarray}
\label{eq:static_aw}
v = [v_1^T \ v_2^T]^T = D_{aw} (u - \sat(u)).
\end{eqnarray}
This signal can be injected into the right hand side of the controller dynamics (\ref{eq:C}) to recover stability and performance of the unconstrained closed-loop system.

When lumping together the plant-controller-anti-windup components (\ref{eq:P}), (\ref{eq:C}), (\ref{eq:static_aw}), $\sigma=\sat(u)$, one obtains the so-called {\em anti-windup closed-loop system}, a nonlinear control system which can be compactly written using the state $x = [x_p^T \ x_c^T]^T$ as in (\ref{eq:CL}) (at the top of the next page),
\begin{figure*}
\begin{eqnarray}
\label{eq:CL}
\begin{array}{rcl}
\dot{x}        & =  &  A_{cl}(q) x   +  \left(B_{cl,q}(q)+B_{cl,v}(q)D_{aw}\right)\dz(u)    +  B_{cl,w}(q)w \\
    z          & =  &  C_{cl,z}(q) x +  \left(D_{cl,zq}(q)+D_{cl,zv}(q)D_{aw}\right)\dz(u)  +  D_{cl,zw}(q)w \\
    u          & =  &  C_{cl,u}(q) x +  \left(D_{cl,uq}(q)+D_{cl,uv}(q)D_{aw}\right)\dz(u)  +  D_{cl,uw}(q)w
\end{array}
\end{eqnarray}
\begin{center}
\line(1,0){500}
\end{center}
\begin{subequations}
\label{eq:analysisALL}
\begin{align}
\label{eq:analysis1}
    &  Q =Q^T>  0, \; U  > 0 \mbox{ diagonal},      \\  
\label{eq:an_bigLMI}
       &   \He\begin{bmatrix}
              A_{cl}(q)Q        &      \left(B_{cl,q}(q)+B_{cl,v}(q)D_{aw}\right)U+  Y^T            &   B_{cl,w}(q) & 0          \\
	              C_{cl,u}(q)Q    &     \left(D_{cl,uq}(q)+D_{cl,uv}(q)D_{aw}\right)U-U         &  D_{cl,uw}(q) & 0         \\
	                    0            &      0               &   -I/2 & 0\\
	              C_{cl,z}(q)Q    &      \left(D_{cl,zq}(q)+D_{cl,zv}(q)D_{aw}\right)U           &    D_{cl,zw}(q)   & -\gamma^2 I/2
	\end{bmatrix} < 0, \quad \textrm{where } \He(Z)=Z+Z^T,\\	
&        \label{eq:analysis2}      \begin{bmatrix}
              Q       &      Y_{[k]}^T          \\
	    Y_{[k]}            &     \bar{u}_k^2/s^2
	\end{bmatrix} \geq 0, \quad k=1,\ldots, n_u, \quad \textrm{where } Y_{[k]} \textrm{ denotes the } k \textrm{-th row of matrix } Y
\end{align}
\end{subequations}
\begin{center}
\line(1,0){500}
\end{center}
\begin{equation}
\label{eq:synthesis}
\begin{array}{rl}
& \hspa \He\!\! \begin{bmatrix}
              A_{cl}(q)Q        &      B_{cl,q}(q)U +B_{cl,v}(q)X +  Y^T      &   B_{cl,w}(q) 		& 0          \\
						C_{cl,u}(q)Q\!\!    &     D_{cl,uq}(q) U +D_{cl,uv}(q)X -U        &  D_{cl,uw}(q) 		& 0         \\
								0            &      								0	               &   -I/2 				& 0\\
				C_{cl,z}(q)Q \! \!      &      D_{cl,zq}(q)U +D_{cl,zv}(q)X           &    D_{cl,zw}(q)   &\!\!\! -\frac{\gamma^2}{2}I
	\end{bmatrix}\! < \! 0,
\end{array}
\end{equation}
\begin{center}
\line(1,0){500}
\end{center}
\end{figure*}
where $\dz$ denotes the deadzone function, \textit{i.e.}, $\dz(u) = u- \sat(u)$, and all the matrices are uniquely determined by the data in (\ref{eq:P}), (\ref{eq:C}), (\ref{eq:static_aw}) (see, \textit{e.g.}, the full authority anti-windup section in \cite{ZackAWbook11} for explicit expressions of these matrices).

The compact form in (\ref{eq:CL}) may be used to represent both the saturated closed loop before anti-windup compensation,  by selecting $D_{aw}=0$,  or the closed loop with anti-windup compensation,  by performing some nonzero selection of~$D_{aw}$. 

We start by analyzing the  system (\ref{eq:CL}) for the \textit{nominal case}, that is when no uncertainty is present and $\Q$ is a singleton
coinciding with the nominal value of the parameters. We recall the following stability and performance analysis result from \cite[Theorem 2]{HuTAC06}.

\begin{proposition}[Regional stability/performance analysis]
\label{prop:AWanalysis}
Given a scalar $s>0$, consider the nominal system, that is let $\Q\equiv\{\bar q\}$.
Assume that the LMI problem \eqref{eq:analysisALL} in the variables $\gamma^2$, $Q$, $Y$ and $U$ is feasible. Then:
\begin{enumerate}[(a)]
	\item the nonlinear algebraic loop in (\ref{eq:CL}) is well posed,
	\item the origin of (\ref{eq:CL}) is locally exponentially stable with
region of attraction containing the set $${\mathcal E}((s^2Q)^{-1}) = \{x:\; x^T Q^{-1} x \leq s^2\},$$
 \item for each $w$ satisfying $\|w\|_2\leq s$, the zero state solution to (\ref{eq:CL}) satisfies $\|z\|_2 \leq \hat{\gamma} \|w\|_2$, where the \Ltwo\ gain of the system is given by
	\begin{eqnarray}\label{eq:L2_computation_analysis}
	\hat{\gamma}^2(s) &=& \displaystyle\min_{\left\{\gamma^2,Q,Y,U\right\}}\gamma^2 \\
	 &&\mbox{s.t. (\ref{eq:analysisALL}).} \nonumber
	\end{eqnarray}
\end{enumerate}
\end{proposition}
\smallskip
As suggested in \cite{HuTAC06}, one may use the result of Proposition~\ref{prop:AWanalysis} to compute an estimate of all the nonlinear \Ltwo\ gain curve (see \cite{Megretski96}, namely a function $s\mapsto \hat \gamma(s)$ such that for each $s$ in the feasibility set of (\ref{eq:analysisALL}) and for each $w$ satisfying $\|w\|_2 \leq s$, the
zero state solution to (\ref{eq:CL}) satisfies 
$$
\|z\|_2 \leq \hat \gamma(s) \|w\|_2.
$$
To do so, it is possible to sample the nonlinear gain curve $s \mapsto \hat \gamma(s)$ by selecting suitable positive values $s_1 < \cdots< s_n$ and, for each $k=1,\ldots, n$, solving Problem (\ref{eq:L2_computation_analysis}), after replacing $s= s_k$.
Then, the \Ltwo\ gain curve estimate can be constructed by interpolating the points $(s_k,\hat{\gamma}_k(s_k))$, $k=1,\ldots, n$.

Following the derivations in \cite{TarbouriechTAC05} (which generalize the global results of \cite{MulderAuto01}), one may notice that the product $D_{aw} U$ appears in a linear way in equation (\ref{eq:an_bigLMI}) and, for a fixed value of $s$, the synthesis of a static anti-windup gain minimizing the nonlinear \Ltwo\ gain can be written as a convex optimization problem, as stated next.

\begin{proposition}[Regional stability/performance synthesis]
\label{prop:AWsynthesis}
Given the plant-controller pair (\ref{eq:P}), (\ref{eq:C}), and a scalar $s>0$,
consider the nominal system, that is $\Q\equiv\{\bar q\}$ is a singleton. Assume that the LMI optimization 
	\begin{eqnarray}\label{eq:L2_computation_synthesis}
	\hat{\gamma}^2(s) &=& \displaystyle\min_{\left\{\gamma^2,Q,Y,U,X\right\}}\gamma^2 \\
	 &&\mbox{s.t. }(\ref{eq:analysis1}), (\ref{eq:analysis2}), (\ref{eq:synthesis}) \nonumber
	\end{eqnarray}
is feasible.
Then, selecting the static anti-windup gain as
\begin{equation}
\label{eq:staticAWsel}
D_{aw}=XU^{-1},
\end{equation}
the anti-windup closed-loop system (\ref{eq:P}), (\ref{eq:C}), (\ref{eq:static_aw}), $\sigma=\sat(u)$ or its equivalent representation in (\ref{eq:CL}) satisfies properties (a)-(c) of Proposition~\ref{prop:AWanalysis}.
\end{proposition}
\smallskip


Consider now the \textit{uncertain case}, when the system matrices in \eqref{eq:CL} defining the dynamics of $x$ and $z$ are continuous (possibly nonlinear) functions of the uncertainty $q \in \Q$. Then, the interest is in finding robust solutions to the analysis and design problems discussed before. For instance, in the analysis case, one could search for common certificates $Q,Y,U$ such that $\gamma^{2}$ is minimized over (\ref{eq:analysisALL}) for all $q\in\Q$. This approach is the one pursued in the paper \cite{ForSaZacDab:13}, where scenario results are used to find probabilistic guaranteed estimates. 

However, as discussed in Section~\ref{sec:SwC}, an approach based on common certificates is in general very conservative, and one would be more interested in finding \textit{parameter-dependent} certificates. To do this, we would need to solve the
following robust optimization problem with certificates
\begin{align}
\hat{\gamma}^2(s) =& \displaystyle\min\gamma^2 \label{eq:RwC-AWA} \\
&\text{s.t. } \gamma^{2}\in
\left\{\gamma^{2} | \ \exists 
\{Q,Y,U\}
\text{ satisfying }\eqref{eq:analysisALL}
\right\} \,\forall q\in\Q.\nonumber
\end{align}
A similar rationale can be applied to robustify the anti-windup synthesis problem of Proposition~\ref{prop:AWsynthesis}. As a matter of fact, when the system matrices are uncertain, one meets similar obstructions to those highlighted as far as analysis was concerned. Again, instead of looking for common Lyapunov certificates as done in \cite{ForSaZacDab:13}, we write the following RwC problem
\begin{align}
\hat{\gamma}^2(s) =& \displaystyle\min\gamma^2 &\label{eq:RwC-AWS} \\
&\text{s.t. } \{\gamma^{2},U,X\}\in \bigl\{ \{\gamma^{2},U,X\}\  | \ \exists 
\{Q,Y\}
\text{ satisfying } \nonumber \\ & (\ref{eq:analysis1}), (\ref{eq:analysis2}), (\ref{eq:synthesis}) \bigr\} \quad\forall q\in\Q.\nonumber
\end{align}

Note that both problems \eqref{eq:RwC-AWA} and \eqref{eq:RwC-AWS} are difficult nonconvex semi-infinite optimization problems, due to the fact that one has to determine the certificates as functions of the uncertain parameter $q$. A classicall approach in this case is to assume a specific dependence (\textit{e.g.}, affine) of the certificates on the uncertainty.
Instead, in this paper we adopt  a probabilistic approach, assuming that $q$ is a random variable with given probability distribution over $\Q$, and apply the SwC approach discussed in Section~\ref{sec:SwC}. This allows us to find an implicit dependence on $q$ of the certificates. This is in the spirit of the original idea proposed in \cite{Oishi:07a}. 

The following two theorems, whose proofs come straightforwardly from Propositions \ref{prop:scenario} and \ref{prop:AWanalysis}, exploit the SwC approach to address the robust nonlinear \Ltwo\ gain estimation  and synthesis for saturated systems. In particular, the next theorem provides a convex optimization procedure to obtain probabilistic information about the worst case nonlinear \Ltwo\ gain.

\smallskip

\begin{theorem}[Probabilistic performance analysis]
\label{thm:analysis}
Fix a positive value $s$ denoting an upper bound for $\left\|w\right\|_2$ and define an acceptable level of probability of constraint violation $\epsilon \in (0,1)$ and a level of confidence $\delta \in (0,1)$. 
Let the design variable and the certificates be, respectively, $\theta=\gamma^2$ and $\xi=\left\{ Q, U , Y \right\}$ and choose $N$ 
as the smallest integer such that  $\B(N,\epsilon,n_{\theta})\leq\delta$.
Construct the SwC approximation of problem \eqref{eq:RwC-AWA} assuming that, for any multisample extraction, the ensuing problem \eqref{eq:scenario_cert} is feasible and attains a unique optimal solution.
Then, for each $\left\|w\right\|_2<s$, the zero state solution of system \eqref{eq:CL} satisfies 
\[
\Pr(\left\|z\right\|_2 > \hat{\gamma}(s) \left\|w\right\|_2)<\epsilon,
\] 
with level of confidence no smaller than $1-\delta$.
\end{theorem}

\smallskip

The next theorem provides the formal result for robust randomized synthesis using the SwC approach. 
\smallskip

\begin{theorem}[Probabilistic anti-windup synthesis]
\label{thm:synthesis}
Fix a positive value $s$ denoting an upper bound for $\left\|w\right\|_2$, and define an acceptable level of probability of constraint violation $\epsilon \in (0,1)$ and a level of confidence $\delta \in (0,1)$. 
Let the design variable and the certificates be, respectively, $\theta=\left\{\gamma^2,X,U\right\}$ and $\xi=\left\{ Q, Y \right\}$ and choose $N$ 
as the smallest integer such that  $\B(N,\epsilon,n_{\theta})\leq\delta$.
Construct the SwC approximation of problem \eqref{eq:RwC-AWS} assuming that, for any multisample extraction, the ensuing problem \eqref{eq:scenario_cert} is feasible and attains a unique optimal solution.
Then, for each $\left\|w\right\|_2<s$, the zero state solution of the uncertain system \eqref{eq:CL} with anti-windup static compensator \eqref{eq:staticAWsel} satisfies
\[
\Pr(\left\|z\right\|_2 > \hat{\gamma}(s) \left\|w\right\|_2)<\epsilon,
\] 
with level of confidence no smaller than $1-\delta$.
\end{theorem}

For completeness, in \eqref{eq:synthesis-SwC} we report the SwC problem arising from the application of Theorem~\ref{thm:synthesis}. The SwC problem arising from the application of Theorem~\ref{thm:analysis} can be constructed following the same rationale and it is not reported here for the sake of space.

The reformulation of the SwC approach for nonlinear gain analysis and synthesis in Theorems~\ref{thm:analysis} and \ref{thm:synthesis} is appealing from an engineering viewpoint. As a matter of fact, since the $N$ instances of the system matrices are extracted according to the probability distribution of the uncertainty, the solution of the problem provides a view of what \textit{may} happen in most of practical situations. Moreover, we stress that the proposed formulation does not constrain the unknown Lyapunov matrices $Q_{i}$'s to be the same for all the sampled perturbations. Instead, it allows them to vary from one sample to another. This is possible because the $Q_{i}$'s (as well as the $Y_{i}$'s) are only instrumental for the computation of the robust compensator.
Notice that, unlike system matrices, \textit{e.g.}, the $A_{cl}(q^{(i)})$'s, which are uncertain by definition, the certificates are unknown but {\em they are not random variables}. 

\begin{figure*}
\begin{equation}
\label{eq:synthesis-SwC}
\begin{array}{rl}
 \hspa \hat{\gamma}^2(s) =& \min\nolimits\limits_{\left\{\gamma^2,Q_{1},\ldots,Q_{N},Y_{1},\ldots,Y_{N},U,X\right\}}\gamma^2 \\
& \hspa \mbox{s.t. } 
  Q_{i} =Q_{i}^T>  0, \; U  > 0 \mbox{ diagonal},      \\  
& \quad \hspa \He\!\! \begin{bmatrix}
              A_{cl}(q^{(i)})Q_i        &      B_{cl,q}(q^{(i)})U +B_{cl,v}(q^{(i)})X +  Y_i^T      &   B_{cl,w}(q^{(i)}) 		& 0          \\
						C_{cl,u}(q^{(i)})Q_i\!\!    &     D_{cl,uq}(q^{(i)}) U +D_{cl,uv}(q^{(i)})X -U        &  D_{cl,uw}(q^{(i)}) 		& 0         \\
								0            &      								0	               &   -I/2 				& 0\\
				C_{cl,z}(q^{(i)})Q_i \! \!      &      D_{cl,zq}(q^{(i)})U +D_{cl,zv}(q^{(i)})X           &    D_{cl,zw}(q^{(i)})   &\!\!\! -\frac{\gamma^2}{2}I
	\end{bmatrix}\! < \! 0\\
&          \begin{bmatrix}
              Q_{i}       &      Y_{i,[k]}^T          \\
	    Y_{i,[k]}            &     \bar{u}_k^2/s^2
	\end{bmatrix} \geq 0, \quad k=1,\ldots, n_u, \quad i=1,\ldots,N
\end{array}
\end{equation}
\begin{center}
\line(1,0){500}
\end{center}
\end{figure*}

\section{SIMULATION EXAMPLE}\label{ref:example}
\begin{figure}[hbt]
\begin{center}
\setlength{\unitlength}{1547sp}%
\begingroup\makeatletter\ifx\SetFigFont\undefined%
\gdef\SetFigFont#1#2#3#4#5{%
  \reset@font\fontsize{#1}{#2pt}%
  \fontfamily{#3}\fontseries{#4}\fontshape{#5}%
  \selectfont}%
\fi\endgroup%
\begin{picture}(9687,2799)(1864,-4348)
\thinlines
\put(1876,-2161){\line( 1, 0){825}}
\put(2701,-2311){\framebox(900,300){}}
\put(3601,-2161){\line( 1, 0){1425}}
\put(4351,-2161){\line( 0,-1){375}}
\put(4201,-3436){\framebox(300,900){}}
\put(4351,-3436){\line( 0,-1){300}}
\put(4051,-3811){\framebox(600,75){}}
\put(4051,-3961){\framebox(600,75){}}
\put(4351,-3961){\line( 0,-1){375}}
\put(5926,-2161){\line( 1, 0){1425}}
\put(6676,-2161){\line( 0,-1){375}}
\put(6526,-3436){\framebox(300,900){}}
\put(6676,-3436){\line( 0,-1){300}}
\put(6376,-3811){\framebox(600,75){}}
\put(6376,-3961){\framebox(600,75){}}
\put(6676,-3961){\line( 0,-1){375}}
\put(5026,-2311){\framebox(900,300){}}
\put(7351,-2311){\framebox(900,300){}}
\put(8251,-2161){\line( 1, 0){1425}}
\put(9001,-2161){\line( 0,-1){750}}
\put(8701,-2986){\framebox(600,75){}}
\put(8701,-3136){\framebox(600,75){}}
\put(9001,-3136){\line( 0,-1){1200}}
\put(9676,-1561){\line( 0,-1){1200}}
\put(9676,-2761){\line( 2, 1){1140}}
\put(10801,-2161){\line(-2, 1){1140}}
\put(10801,-2161){\line( 1, 0){450}}
\put(1876,-4336){\line( 1, 0){9375}}
\put(2026,-4111){\vector( 0, 1){1725}}
\put(11026,-4111){\vector( 0, 1){1725}}
\put(2251,-3511){\makebox(0,0)[lb]{\smash{\SetFigFont{13}{34.8}{\rmdefault}{\mddefault}{\itdefault}$V_{i}$}}}
\put(11326,-3436){\makebox(0,0)[lb]{\smash{\SetFigFont{13}{34.8}{\rmdefault}{\mddefault}{\itdefault}$V_{o}$}}}
\put(5651,-4036){\makebox(0,0)[lb]{\smash{\SetFigFont{13}{34.8}{\rmdefault}{\mddefault}{\itdefault}$C_{2}$}}}
\put(7976,-3211){\makebox(0,0)[lb]{\smash{\SetFigFont{13}{34.8}{\rmdefault}{\mddefault}{\itdefault}$C_{3}$}}}
\put(3326,-4036){\makebox(0,0)[lb]{\smash{\SetFigFont{13}{34.8}{\rmdefault}{\mddefault}{\itdefault}$C_{1}$}}}
\put(3401,-3211){\makebox(0,0)[lb]{\smash{\SetFigFont{13}{34.8}{\rmdefault}{\mddefault}{\itdefault}$R_{2}$}}}
\put(5726,-3211){\makebox(0,0)[lb]{\smash{\SetFigFont{13}{34.8}{\rmdefault}{\mddefault}{\itdefault}$R_{4}$}}}
\put(9976,-2311){\makebox(0,0)[lb]{\smash{\SetFigFont{13}{34.8}{\rmdefault}{\mddefault}{\itdefault}k}}}
\put(7651,-1861){\makebox(0,0)[lb]{\smash{\SetFigFont{13}{34.8}{\rmdefault}{\mddefault}{\itdefault}$R_{5}$}}}
\put(5326,-1861){\makebox(0,0)[lb]{\smash{\SetFigFont{13}{34.8}{\rmdefault}{\mddefault}{\itdefault}$R_{3}$}}}
\put(3001,-1861){\makebox(0,0)[lb]{\smash{\SetFigFont{13}{34.8}{\rmdefault}{\mddefault}{\itdefault}$R_{1}$}}}
\end{picture}
\caption{\label{fig:network} The passive electrical network with saturated input voltage.}
\end{center}
\end{figure}
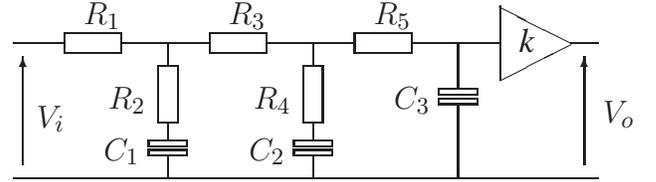

In this section, we show the effectiveness of the proposed approach, by designing an anti-windup compensator for the passive electrical network in Fig. \ref{fig:network}. The circuit is a benchmark example in the anti-windup literature and was already employed in \cite{ZackAWbook11} to show the potential of static compensators.

The dynamics of the network is determined by 5 resistors and 3 capacitors, whose nominal values are reported in Table \ref{tab:parameters}. The gain $k$ is instead selected such that the transfer function between $V_i$ and $V_o$ is monic.
\begin{table}
\centering
\begin{tabular}{|c||c|c|}
\hline
\textbf{name}&\textbf{value}&\textbf{units}\\
\hline
$R_1$ & $313$  &$\mathrm{\Omega}$\\ 
$R_2$ & $20$   &$\mathrm{\Omega}$\\ 
$R_3$ & $315$  &$\mathrm{\Omega}$\\ 
$R_4$ & $17$   &$\mathrm{\Omega}$\\
$R_5$ & $10$   &$\mathrm{\Omega}$\\ 
$C_1$ & $0.01$ &$\mathrm{F}$\\
$C_2$ & $0.01$ &$\mathrm{F}$\\ 
$C_3$ & $0.01$ &$\mathrm{F}$\\
\hline
\end{tabular}
\caption{Nominal parameter values for the network in Fig. \ref{fig:network}.}
\label{tab:parameters}
\end{table}

The nominal plant is then of the form \eqref{eq:P}, where:
\begin{align}
\footnotesize
\left[
\begin{array}{c|c|c}
A_p     & B_{p,u}   & B_{p,w}\\    \hline
C_{p,z} & D_{p,zu}  &   D_{p,zw}\\ \hline
C_{p,y} & D_{p,yu}  &   D_{p,yw}
\end{array}
\right] =
\left[
\begin{array}{cc|c|c|c}
 -10.6  &  -6.09  & -0.9 & 1 &0\\ 
    1   &    0    &   0  & 0 &0\\
    0   &    1    &   0  & 0 &0\\ \hline
   -1   &   -11   &  -30 & 0 &0\\ \hline
    1   &   11    &  30 & 0 &0\\
\end{array}
\right], \nonumber
\normalsize
\end{align}
$w$ represents the reference value for the output voltage $V_o$ and $z=w-y$.

The controller is a PID and it is designed based on the nominal model, such that the nominal phase margin is $89.5$ degrees and the nominal gain margin is infinite. In the form \eqref{eq:C}, the controller is expressed by the matrices:
\begin{align}
&\left[
\begin{array}{c|c|c}
A_c   & B_{c,y} & B_{c,w} \\ \hline
C_{c} & D_{c,y} & D_{c,w}
\end{array}
\right] =
\left[
\begin{array}{cc|c|c}
-80   & 0    & 1  &-1\\ 
1     & 0    & 0  &0\\ \hline
20.25 & 1600 & 80 &-80\\ 
\end{array}
\right].\nonumber
\end{align}
Assume now that the input $V_i$ is saturated between the minimum and the maximum voltage $\pm 1$. An anti-windup compensator based on the nominal model can be designed to minimize the nonlinear gain between the reference and the tracking error. The gain obtained using $s=0.003$ is $D^{nom}_{aw}=[-0.0855,0.0011,0.9887]^T$. 

Under the hypothesis that the parameters are Gaussian distributed with mean values as in Table I and standard deviation of $10\%$, also the robust randomized compensator can be computed. Using the same value of $s$ of the nominal case, \textit{i.e.}, $s=0.003$, $D^{rand}_{aw}=[-2.1493,0.0266,0.6407]^T$ is obtained. For this example, $\epsilon=0.01$ and $\delta=10^{-6}$ are selected. Therefore, we notice that, in principle, at least $N=7565$ are required to give the desired probabilistic guarantees ($n_{\theta}=35$ in \eqref{eq:binom2}). However, since Algorithm 1 is employed, only $2270$ samples are actually used for the design.

\begin{figure}[h!]
\centering
\includegraphics[width = 1\columnwidth]{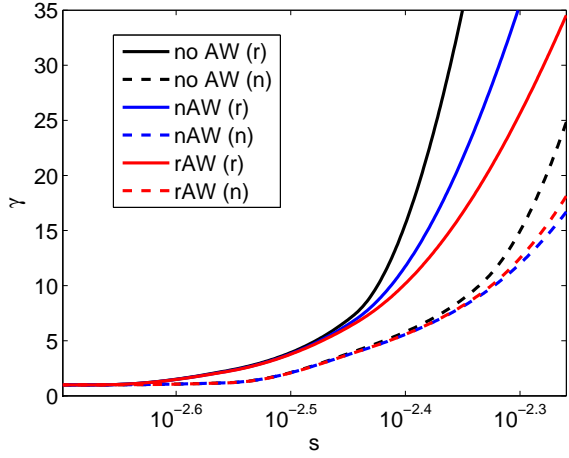}
\caption{\Ltwo\ gain estimates for the nonlinear closed-loop systems with and without anti-windup compensator. Both robust (solid) and nominal (dashed) analysis are considered to assess the performance of robust (blue) and nominal (red) compensators with respect to the system without anti-windup augmentation (black).}
\label{fig:L2gain}
\end{figure}

The results of nominal and robust anti-windup compensators are reported in Fig. \ref{fig:L2gain}, where both the nominal and the robust \Ltwo\ gain estimates are computed.
As expected, we observe that the robust compensator outperforms the one designed for the nominal system, as far as the robust \Ltwo\ gain is concerned (solid lines). Conversely, when the performance is evaluated on the nominal system (dashed lines), the robust compensator yields worse results, since it is more conservative. From the analysis point of view, notice also that the \Ltwo\ gain estimated using the robust probabilistic method is larger than the one given by nominal analysis. This holds for any configuration of the saturated closed-loop system (without anti-windup, with nominal compensator and with robust compensator).

\begin{figure}[h!]
\centering
\includegraphics[width = 1\columnwidth]{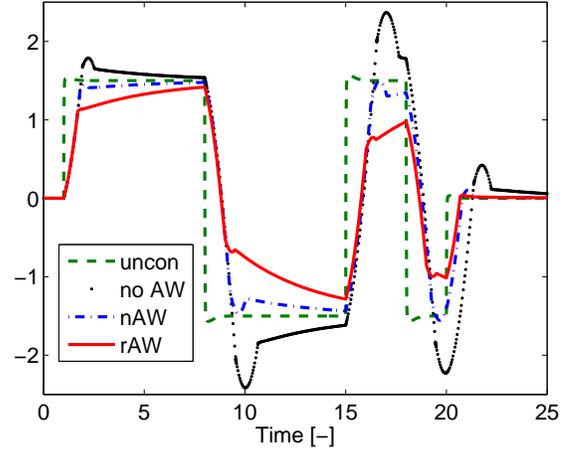}
\caption{Time responses of the closed-loop system with nominal parameters and different configurations of the anti-windup architecture: unconstrained system (dashed), saturated system without anti-windup compensator (dotted), saturated system with nominal anti-windup (dash-dotted) and saturated system with robust anti-windup (solid).}
\label{fig:time_nominal}
\end{figure}

\begin{figure}[h!]
\centering
\includegraphics[width = 1\columnwidth]{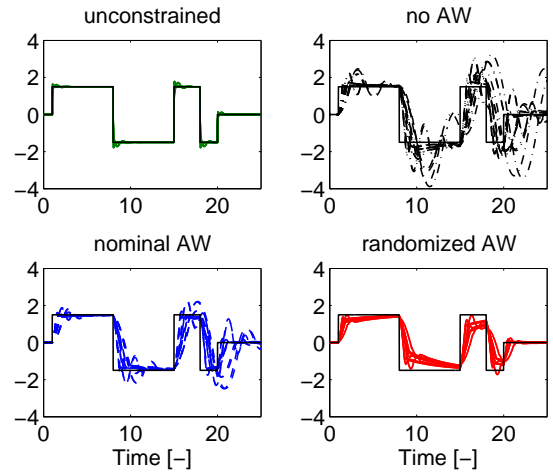}
\caption{Time responses of the uncertain closed-loop system with different configurations of the anti-windup architecture.}
\label{fig:time_robust}
\end{figure}

The degradation of the performance of the nominal closed-loop system using the robust compensator in place of the nominal one can be assessed by looking at the time responses illustrated in Fig. \ref{fig:time_nominal}. From the figure, we conclude that, although in any case the use of a compensator increases the overall performance in terms of tracking error and overshoot, we have to accept worse behavior in nominal conditions when using robust anti-windup. 
However, this choice is rewarding when acting on a system subject to uncertainty. Specifically, in Fig. \ref{fig:time_robust}, we show that the response of the system with the robust anti-windup is less sensitive to parameter uncertainties than the other solutions. Specifically, in the figure, we illustrate, as an example, the responses corresponding to $16$ different combinations of $\pm 10 \%$ perturbation of the nominal parameters.

\section{CONCLUSIONS}
In this paper, we proposed a distinction between standard robust convex optimization problems (RO), where all the optimization variables are used for design, and robust optimization problems with certificates (RwC), where some optimization variables play just the role of certificates of performance. Then, we defined the scenario counterpart for randomized solution of RwC, that we called Scenario with Certificates (SwC). Unlike standard scenario optimization (SO), SwC allows one to create one certificate variable for each sample of the uncertainty, thus leading to less conservative solutions.
Static anti-windup augmentation is a typical problem where some variables, like the Lyapunov matrix, appear in the LMIs only to certify stability of the closed-loop system, but are not used inside the formula to compute the compensator gains. We applied the SwC approach to anti-windup compensator design and showed that this method leads to better design than the state of the art.
In future works, we will concentrate on the application of such an approach on a broader class of problems, to better analyze the potential of SwC with respect to standard scenario optimization.


\bibliographystyle{plain}
\bibliography{randomized_methods,saturate,aw,aw2,zack,Frugi-biblio}

\end{document}